# Current-induced successive structural phase transitions beyond thermal equilibrium in single-crystal VO$_2$

Shunsuke Kitou[1]*, Akitoshi Nakano[2]*, Masato Imaizumi[2], Yuiga Nakamura[3],

Yuto Nakamura[4], Hideo Kishida[4], Taka-hisa Arima[1,5], Ichiro Terasaki[2]

[1]*Department of Advanced Materials Science, The University of Tokyo, Kashiwa 277-8561, Japan*

[2]*Department of Physics, Nagoya University, Nagoya 464-8602, Japan*

[3]*Japan Synchrotron Radiation Research Institute (JASRI), SPring-8, Hyogo 679-5198, Japan*

[4]*Department of Applied Physics, Graduate School of Engineering, Nagoya University, Nagoya 464-8603, Japan*

[5]*RIKEN Center for Emergent Matter Science (CEMS), Wako 351-0198, Japan*

*Corresponding authors.

## Abstract

Nonequilibrium systems driven by external energy sources host unexplored physics; yet phase transitions beyond thermal equilibrium remain elusive. Here, we demonstrate that electric current induces structural phase transitions in single-crystal VO$_2$, a prototypical strongly correlated material. At room temperature, synchrotron X-ray diffraction shows that a current density of 6.5 A/cm$^2$ disrupts V-V dimers, driving a monoclinic-to-tetragonal insulator-to-metal transition, independent of Joule heating. Increasing the current to 10 A/cm$^2$ triggers a discontinuous isotropic lattice expansion, stabilizing a novel tetragonal structure that does not exist in thermal equilibrium. Optical microscopy and microscopic Raman spectroscopy reveal dynamic domain motion and metastable phases, reminiscent of dissipative structures. These findings establish direct pathways to access hidden phases and symmetry changes beyond thermal equilibrium, broadening the frontiers of nonequilibrium thermodynamics.

**Main text**

Nonequilibrium systems that continuously exchange energy and matter with their surroundings can give rise to striking macroscopic order, sustained by persistent energy fluxes and entropy production. From the coordinated motion of bird flocks to planetary-scale atmospheric vortices, such dissipative structures manifest across a remarkable range of natural phenomena. In crystalline solids, atomic ordering during growth exemplifies self-assembly governed by thermodynamic principles. While solid-state physics has successfully embedded these concepts within equilibrium frameworks—providing profound insights into emergent quantum phases—the exploration of nonequilibrium phenomena in solids, while increasingly attracting attention [1], remains in its infancy. A central question in nonequilibrium thermodynamics is whether crystalline solids, when driven far from thermal equilibrium, can undergo transitions to nonequilibrium phases as steady states, accompanied by symmetry breaking and the emergence of dynamic dissipative structures. Electric current offers a direct route to drive such nonequilibrium phases, where nonlinear responses serve as key signatures of departure from equilibrium. Yet, although nonlinear conductivity and associated structural changes has been extensively studied in various materials, including organic salts [2,3] and transition metal oxides [4-9], clear experimental evidence of current-induced structural phase transitions beyond thermal effects has remained elusive.

Vanadium dioxide ($VO_2$) is a prototypical strongly correlated electron system, renowned for its insulator-to-metal transition near 340 K upon heating [10-17], where the breaking of V-V dimers drives a symmetry change from a monoclinic to a tetragonal structures. The mechanism of the insulator-to-metal transition remains controversial, involving molecular orbital formation [18-22], electron correlation [23-25], and electron-lattice interaction [26-28]. Despite its simple composition and structure, $VO_2$ exhibits diverse electronic properties under chemical substitution [29,30] and external fields [31-41]. At room temperature, dramatic structural and electronic changes occur under light irradiation [31-35] and high magnetic fields [36], suggesting the disappearance of V-V dimers. However, these effects manifest in transient excited states rather than in steady states. Another intriguing phenomenon is nonlinear conduction under applied electric fields [37-39] and currents [40,41]. Previous studies have reported an electric-field-induced monoclinic-to-tetragonal transition in single-crystal [37], thin-film [39,42-44], and nanobeam samples [45,46]. However, since these transitions are driven by strong electric fields ($E \gg 10$ kV/cm), it was difficult to rule out the effect of Joule heating, raising doubts about whether the transitions are genuine nonequilibrium phase transitions. Recent studies employing direct current have unveiled a distinct insulator-to-metal transition in single-crystal $VO_2$ [41], revealing a novel metallic state fundamentally different from Joule-heating-induced metallic phase under an electric field [37-39,42-46]. Notably, because Joule heating diminishes as resistance decreases under constant current, this approach mitigates thermal artifacts and opens a new avenue to explore genuine nonequilibrium phase transitions. A crucial question thus arises: does this current-induced transition involve an intrinsic change in crystal symmetry, signaling the emergence of a truly nonequilibrium structural phase?

In this paper, we report current-induced structural phase transitions beyond thermal equilibrium in single-crystal $VO_2$, observed via X-ray diffraction (XRD), optical microscopy, and microscopic Raman spectroscopy. Our findings demonstrate that these transitions occur independently of Joule heating and shed light on the intrinsic structure of nonequilibrium phases in crystalline solids. Figure 1 shows the current density versus temperature ($J$-$T_0$) phase diagram and schematic representations of the crystal structure in each phase, obtained from our synchrotron XRD experiments. Here, $T_0$ indicates the temperature at which the current is zero. We identified two new phases, shown in the pink and green regions, in areas where current was applied. Figures 2A and 2B show a schematic of X-ray diffractometer and a photograph of a $VO_2$ single crystal used in the current-induced XRD experiments, respectively. Figures 2C-2G show the XRD data at various temperatures $T_0$ and current densities $J$. A schematic of the reciprocal lattice points of the M1 and R lattices, corresponding to the XRD data, is shown in Figure 2H. At $T_0 = 300$ K and $J = 0$ A/cm$^2$, there are several Bragg peaks attributed to the M1 phase in Fig. 2E. As the temperature increases, at $T_0 = 345$ K and $J = 0$ A/cm$^2$, the number of Bragg peaks is reduced by half (Fig. 2G) compared to the M1 phase. This reduction indicates a monoclinic-to-tetragonal transition from the M1 to high-$T$ R structures, accompanied by a halving of the unit cell volume. It is noted that weak diffuse scattering intensity is observed along the $\boldsymbol{a}_t^* + \boldsymbol{c}_t^*$ direction in the high-$T$ R structure (Fig. 2G and Fig. S7A), corresponding to the short-range order of the V displacements [22].

By applying current along the $c_t$-axis at $T_0 = 300$ K, new peaks appear above $J = 6.5$ A/cm$^2$ (Fig. 2D). The newly appearing peaks are located only near the positions of the Bragg peaks of the high-$T$ R structure, with no new peaks observed near those emerging from the M1 structure. This result suggests that the new phase induced by current application has the same symmetry as the high-$T$ R structure, resulting in the disappearance of V-V dimers and an insulator-to-metal transition. In this $J$ region, monoclinic and tetragonal structures coexist as domains. We designate this R structure under applied current as the low-$J$ R structure. In the low-$J$ R structure, no strong diffuse scattering was observed, unlike in the high-$T$ R structure (see Fig. S7), suggesting an absence of short-range order among V atoms. Therefore, despite sharing the same lattice symmetry, the high-$T$ R and low-$J$ R structures appear to possess fundamentally different structures. The effect of Joule heating due to the application of current was investigated using a thermographic camera at $T_0 = 300$ K (Fig. S1). The maximum temperature of the sample increased by approximately 325 K, which is well below the phase transition temperature $T_C = 340$ K. Based on these experimental findings, we conclude that the current-induced phase transition is not a result of Joule heating, which aligns with the previous nonlinear conduction measurements [41].

When a higher current density of $J \geq 10$ A/cm$^2$ was applied, the peaks corresponding to the low-$J$ R structure disappeared, and new peaks emerged (Fig. 2C and Fig. S5), which also has the same tetragonal symmetry as the high-$T$ R and low-$J$ R structures. Consequently, we refer to this R structure in the $J \geq 10$ A/cm$^2$ region as the high-$J$ R. This phase transition at higher current densities was not observed in nonlinear conduction measurements [41], as the resistance was sufficiently low at these high current densities. At $T_0 = 345$ K and $J = 12$ A/cm$^2$, the Bragg peaks corresponding to the M1, low-$J$ R, and high-$J$ R structures disappear (Fig. 2F), resulting in a pattern almost identical to that of the single high-$T$ R structure (Fig. 2G).

The observed change in the peak positions from the low-$J$ R to the high-$J$ R structures at $T_0 = 300$ K suggests discontinuous changes in the lattice parameters. From the peak positions of the tetragonal lattice at each XRD data, we investigate the current density dependence of the lattice parameters at $T_0 = 300$ K, as shown in Figures 3A-3D.

The $a_t$ and $c_t$ parameters of the low-$J$ R structure (black circles in the $J$ = 6.5—9 A/cm$^2$ range) are comparable to those in the high-$T$ R structure at $T_0$ = 345 K (red squares), as shown in Figs. 3A and 3B. In contrast, the $a_t$ and $c_t$ parameters of the high-$J$ R- structure (black circles in the $J \geq 10$ A/cm$^2$ range) are larger than those in both the high-$T$ R and low-$J$ R structures. Furthermore, the cell volume $V_t$ increases sharply by 0.28(14)% during the phase transition from the low-$J$ R to high-$J$ R structures around $J$ = 10 A/cm$^2$ (Fig. 3C), while the $a_t/c_t$ ratio remains nearly unchanged (Fig. 3D), indicating isotropic volume expansion.

We compare the changes of the tetragonal lattice parameters induced by current application and temperature increase. Figures 3E and 3F show the temperature dependence of $V_t$ and $a_t/c_t$ of the high-$T$ R structure without current, respectively. The $V_t$ value increases monotonically with increasing temperature. The volume expansion rate of 0.28(14)%, resulting from the current-induced phase transition from the low-$J$ R to high-$J$ R structures, corresponds to approximately 80 K in terms of thermal expansion with temperature. The $a_t/c_t$ ratio decreases monotonically with increasing temperature, with a significant elongation along the $c_t$-axis (inset of Fig. 3F). This anisotropic lattice expansion due to temperature change (Fig. 1D) is qualitatively different from the isotropic lattice change caused by the current application (Fig. 1B). Therefore, by applying current to VO$_2$ in the insulating phase, we successfully observed two successive structural phase transitions to nonequilibrium low-$J$ and high-$J$ structures, where the V-V dimers are disrupted and no strong short-range order exists among the V atoms (Fig. S7). In particular, the high-$J$ R structure, characterized by an isotropically expanded tetragonal lattice, represents a nonequilibrium phase that cannot exist in thermal equilibrium. This strongly supports the conclusion that the current-induced R structures are not driven by Joule heating.

XRD experiments using a beam diameter of approximately 0.2 mm revealed that the R structure ratio, which appeared under applied current, was approximately 10% of the total domain (Fig. S5). The presence of such a large ratio of nonequilibrium R structure domain excludes the possibility that the nonlinear conductivity in VO$_2$ single crystals arises from the formation of filamentary paths due to severe Joule heating effects, as observed in thin films [38,39,44] and nanobeam samples [45,46]. However, it remains unclear what kind of domains form in single crystals under continuous current flow. To capture the spatial distribution and time evolution of domains in real space, optical micrographs were taken. Figures 4A-4C show the current density dependence of $VO_2$ images at room temperature. It is noted that the crystal used for the optical microscope measurement differs from the one used in the XRD experiments. At $J$ = 0.05 A/cm$^2$, the $VO_2$ crystal appears light green in the image, indicating an insulating phase, as shown in Fig. 4A. As the current density increases, dark green regions emerge above $J$ = 5 A/cm$^2$, confirming the presence of two types of domains—light green and dark green— at $J$ = 7.25 A/cm$^2$ (Fig. 4B). The newly emerged dark green domains likely correspond to the metallic phase. Thus, the light green and dark green domains are assigned to the M1 and low-$J$ R structures, respectively. The typical size of each domain ranges from 10 to 100 μm, which is consistent with the correlation length of approximately 100 μm estimated from nonlinear conduction measurements [41]. This size range is 10 to 100 times larger than what is observed in thin films [38,44] and nanobeam samples [45,46].

Surprisingly, in the $J$ = 5 to 8.5 A/cm$^2$ range—approximately corresponding to the pink region in the phase diagram in Fig. 1C—the dark green domains were observed sliding along the direction of the current (see Movie S1). Around $J$ = 8.5 A/cm$^2$, the domain structure becomes significantly more distorted and finer (Fig. 4C), suggesting a phase transition from the low-$J$ R to high-$J$ R structures, accompanied by a volume expansion. The pattern appears to separate into a large insulating-phase domain and striped domains consisting of metallic and insulating phases. The orientation of the stripe domains appears to be independent of the current direction. In the $J$ = 8.5 to 15 A/cm$^2$ range, no significant changes and motions in the domain structure were observed. Afterward, the current was turned off, and all the crystals reverted to a light green color.

The changes in domain structures under applied currents appear consistent with the emergence of nonequilibrium low-$J$ and high-$J$ R structures detected in XRD experiments. The observed emergence and motion of the domains resemble spontaneous dissipative structures in nonequilibrium systems found in nature. A similar stripe-shaped domain has been observed during the current-induced insulator-to-metal transition in single-crystalline $Ca_2RuO_4$ [6] and K-TCNQ [3], both of which exhibit nonlinear conduction. In $Ca_2RuO_4$, the metallic phase expands throughout the entire sample as the current density increases. In contrast, in $VO_2$, (i) a two-step structural phase transitions occur, and (ii) dissipative-like structures with complex domain patterns emerge, indicating qualitatively different behavior from that of $Ca_2RuO_4$ and K-TCNQ. Although a theoretical framework to fully comprehend this mechanism has yet to be established, these findings may provide a foundation for exploring dissipative structures in crystalline solids.

To investigate the structure of each domain observed from optical micrographs, we performed Raman spectroscopy using a light spot size of 2 μm. Figure 4D shows the position dependence of the Raman spectra under an applied current of 15.00 A/cm². Each spectrum, labeled (i)-(v) in Fig. 4D, corresponds to the positions (i)-(v) shown in Fig. 4C. Note that points of the same color in Fig. 4C exhibit almost identical spectra (see Fig. S8). We observed five distinct spectra: M1, M1', T, M2, and R, with each structure assigned based on previous Raman spectra measurements [30]. In the broad areas colored light green and dark green, spectra corresponding to the insulating M1 and metallic R structures are observed, respectively, which is consistent with the electronic states predicted from the surface colors. It is noteworthy that in the light green area, sandwiched between the dark green areas, the M1', M2, and T structures were observed. Here, the M1' corresponds to the M1 structure where the polarization direction of the light is slightly shifted from $E_i \perp c_t$. The M2 corresponds to another monoclinic structure, distinct from the M1 structure, while the T corresponds to the triclinic structure. The M2 and T structures in $VO_2$ have been identified as resulting from element substitution [29,30] and uniaxial stress [47]. Recent thermopower measurements suggests that the effects of applied current and applied pressure on $VO_2$ are equivalent [48]. In other words, the insulating domains sandwiched between the metallic R structures experience uniaxial stress due to the influence of the surrounding deformed lattice. As a result, in addition to the M1 structure, metastable T and M2 structures may also appear in the light green areas. The energy degeneracy of these multiple metastable structures in $VO_2$ may explain the difference between its domain pattern and the relatively simple stripe-shaped patterns found in $Ca_2RuO_4$ [6] and K-TCNQ [3]. These metastable T and M2 structures were not observed in the XRD experiments, possibly due to the small volume fraction of these domains.

The peculiar structures and dynamics identified in the nonequilibrium phase of $VO_2$

raise a series of fundamental questions, opening new avenues for exploration. First, the observed nonlinear conduction and structural phase transitions occur at remarkably low threshold electric fields ($E_c$), on the order of tens to hundreds of V/cm, which is about three orders of magnitude smaller than K-TCNQ [3] and comparable to those reported for $Ca_2RuO_4$ [4]. Second, the mechanism by which the application of current alone stabilizes the high-$J$ R structure—an isotropically expanded metallic structure—remains elusive. Given that weak $E_c$ is insufficient to directly modulate the electronic state of strongly correlated systems, these findings underscore the pivotal role of current flow itself. Third, the previous nonlinear conduction study in $VO_2$ showed a gradual decrease in the energy gap with increasing current before the transition to the low-$J$ R structure [41]. A similar gap reduction has also been observed in $Ca_2RuO_4$ under current application [4], which is accompanied by a change in the lattice constant [5]. However, no such systematic lattice change is observed in the M1 structure of $VO_2$ (Fig. S6). Fourth, the emergence of a complex coexistence of metallic and insulating domains in the nonequilibrium phase prompts questions about the intrinsic drivers of such inhomogeneity. Finally, while extensive metallic domains spanning 10 to 100 μm are formed under current flow, the detailed distribution of current within these domains remains unclear. Ultimately, the observed structural phase transitions beyond thermal equilibrium represent an unprecedented phenomenon, defying explanation within the conventional framework of nonequilibrium thermodynamics. Continued investigation in this direction promises to deepen understanding of the complex interplay among electronic correlation, lattice dynamics, and external driving forces in nonequilibrium phases.

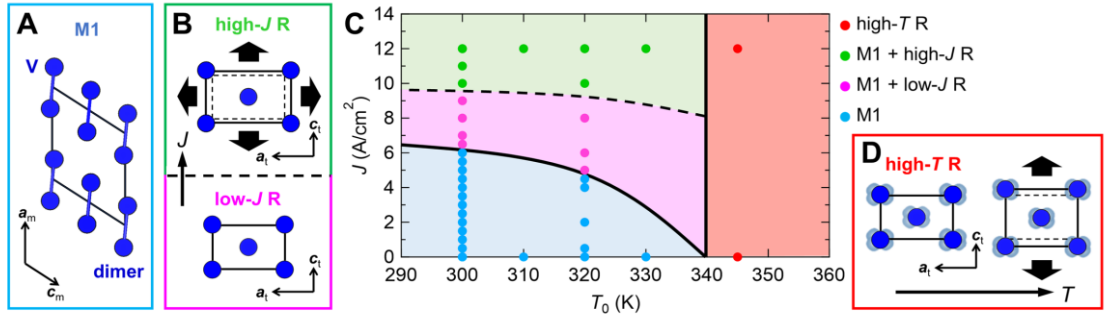

**Fig. 1. Crystal structures of VO$_2$ in the current density versus temperature ($J$-$T_0$) phase diagram.** (**A**) Schematic of the low-temperature M1 structure, in which V atoms form dimers. (**B**) Schematic showing the change in the crystal structure from the low-$J$ R to high-$J$ R structures under applied current. (**C**) $J$-$T_0$ phase diagram identified through single-crystal XRD experiments, where $T_0$ indicates the temperature at which the current is zero. Circles represent measurement points, with red, green, pink, and blue area indicating different phases. (**D**) Schematic showing the change in the crystal structure in the high-$T$ R structure with increasing temperature. V atoms exhibit one-dimensional short-range order along the <111> axes [22]. The unit cell axes of each phase have approximate relationships of $a_m = 2c_t$, $b_m = b_t$, and $c_m = -a_t - c_t$, where the subscripts t and m in the axis labels indicate the tetragonal (R) and monoclinic (M1) lattices, respectively.

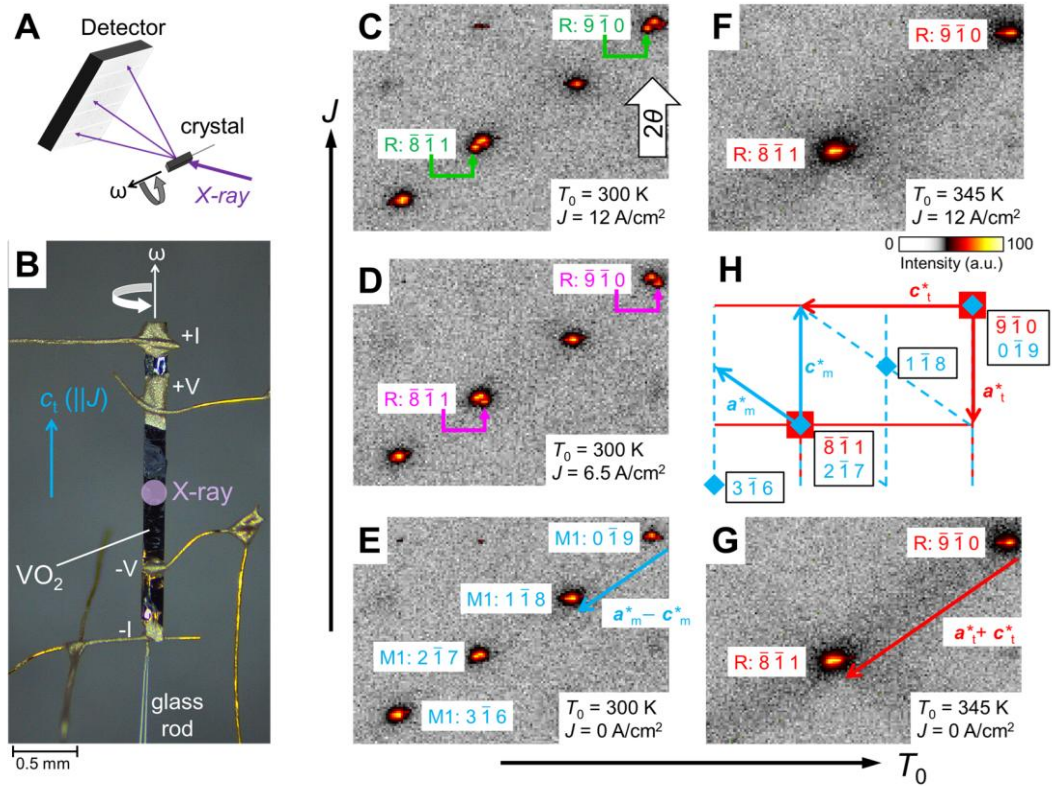

**Fig. 2. XRD experiments under applied current using a VO₂ single crystal.** (**A**) Schematic of XRD measurements. (**B**) A single crystal of VO₂ used for XRD experiments. The crystal is mounted on a glass rod, and a current is applied along the $c_t$-axis using the four-probe method. A light purple circle, representing a beam diameter of 0.2 mm, indicates the position where the X-rays were irradiated. XRD data were collected while rotating the sample around the $\omega$-axis by 180°. (**C-G**) XRD images at various temperatures $T_0$ and current densities $J$. (**H**) Schematic of the reciprocal lattice points of the R and M1 lattices. Red squares and light blue diamonds denote the positions of the Bragg peaks for the R and M1 lattices, respectively.

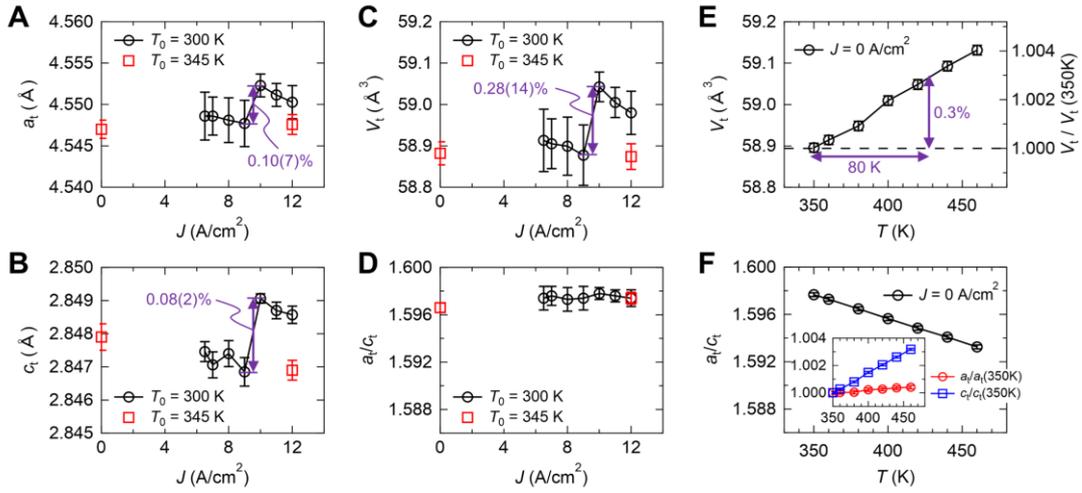

**Fig. 3. Lattice parameters of VO₂.** (**A-D**) Current density dependence of the tetragonal lattice constants $a_t$, $c_t$, volume $V_t$, and the ratio of $a_t/c_t$. (**E,F**) Temperature dependence of the tetragonal lattice parameters $V_t$, and $a_t/c_t$ of the high-$T$ R structure. The inset of (**F**) shows the lattice constants $a_t$ and $c_t$, normalized to 350 K.

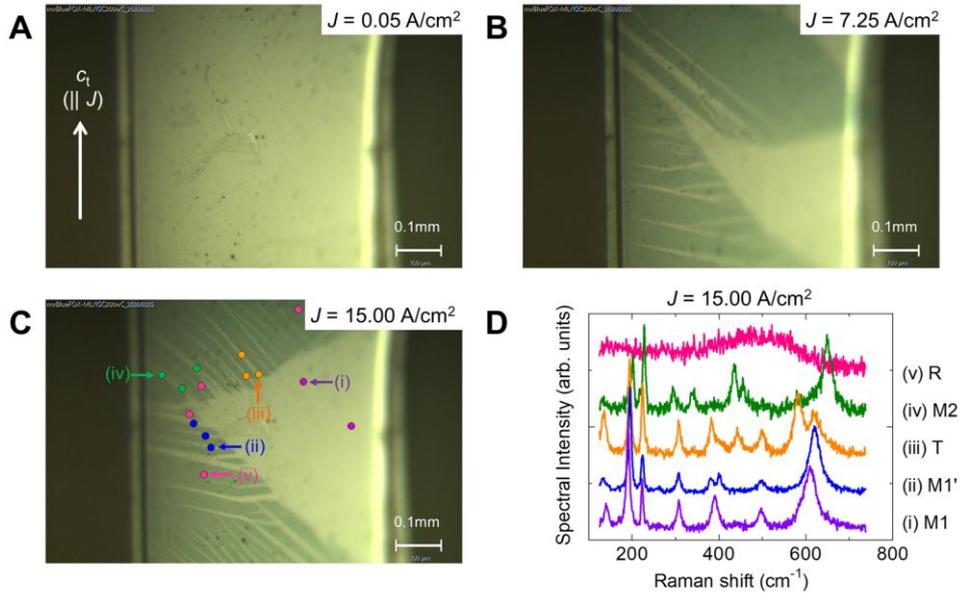

**Fig. 4. Optical microscope images and Raman spectra of a VO₂ single crystal.** (**A-C**) Current density dependence of VO₂ images at room temperature under applied currents of 0.05, 7.25, and 15.00 A/cm². (**D**) Position dependence of the Raman spectra under an applied current of 15.00 A/cm².

## Acknowledgements

We acknowledge T. Oka for the fruitful discussions. **Funding:** This work was supported by JSPS KAKENHI (Grant Nos. 24K17006, 24H01644, and 24K21733), and JST FOREST (Grant No. JPMJFR2362). The synchrotron radiation experiments were performed at SPring-8 with the approval of the Japan Synchrotron Radiation Research Institute (JASRI) (Proposal No. 2023B1603, 2024A1511, 2024A1709, and 2024B1599). **Author contributions:** S.K. and A.N. designed and coordinated the study. M.I. and A.N. grew the single crystals. S.K., A.N., M.I., and Yuiga N. performed the XRD experiment. S.K. analyzed the XRD data. M.I., A.N., S.K., and Yuto N. performed the optical microscopy and microscopic Raman spectroscopy measurements. S.K. and A.N. wrote the manuscript under the supervision of H.K., T.A, and I.T. All authors discussed the experimental results and contributed to the manuscript. **Competing interests:** The authors declare no competing interests. **Data and materials availability:** All data needed to evaluate the conclusions in the paper are present in the paper or the supplementary materials.

# Supplementary Materials

## Materials and Methods

### Sample preparation

Single crystals of $VO_2$ were grown by melting $V_2O_5$ at 950℃ under $N_2$ gas flow conditions (4 L/min) for three days by modifying the growth method in Ref. [49]. The single crystals used in this study were the same as those in Ref. [41].

### Current-induced XRD experiments

XRD experiments were performed on BL02B1 at a synchrotron facility SPring-8, Japan [50]. An $N_2$-gas-blowing device was employed to control the sample temperature. A two-dimensional detector CdTe PILATUS was used to record the diffraction pattern. The X-ray wavelength was $\lambda = 0.30868$ Å. The lattice parameters and crystal orientations were calculated using CrysAlisPro program [51]. The resistivity along the $c_t$ axis was measured using the standard four-probe dc method. Several external currents were applied using a current source (KEITHLEY 6221). Note that we used dc currents instead of dc voltages for the control parameter to avoid thermal runaway associated with nonlinear conduction. To reduce contact resistance, which may be an extrinsic origin of nonlinear conduction due to Joule heating, we deposited a thick gold layer (~100 nm) on the sample surface before bonding the gold wires by using Ag paste.

### Raman microscopy

Raman scattering spectra were obtained using a Renishaw in Via Raman microscope. The excitation light source was a He-Ne laser (wavelength 633 nm) linearly polarized perpendicular to the $c_t$-axis of $VO_2$ and incident on the sample in a backscattering configuration. The spot size of the incident light is approximately 2 μm. The power of the incident light is 40 μW, and the temperature and domain structure were not changed by the incident light.

## Supplementary Text

Figure S1 shows the current density dependence of thermographic images of $VO_2$ at $T_0 = 300$ K. The single crystal used was the same as in the XRD experiments (Fig. 2). The maximum temperature of the sample surface is about 325 K, well below the phase transition temperature $T_C = 340$ K. Figure S2 shows the current density dependence of electric fields and resistivities at $T_0$ = 300 and 320 K. These data were collected simultaneously with the XRD experiments (Fig. 1C). Clear nonlinear conduction is observed, which is consistent with a previous report [41].

Figure S3 shows the whole XRD image of VO$_2$ at $T_0$ = 300 K and $J$ = 0 A/cm$^2$. Figure S4 shows the current density dependence of XRD data at $T_0$ = 300 and 320 K, which is displayed in chronological order. After the measurement at $J$ = 12 A/cm$^2$, the XRD pattern returned to the original single M1 structure once the current was turned off.

Figure S5 shows the current density dependence of the domain ratio of the R structure at $T_0$ = 300 K. The domain ratio is estimated from the intensity ratio of $\boldsymbol{K}_R$ = (-12, 1, 1)$_R$ and $\boldsymbol{K}_{M1}$ = (2, 1, 11)$_{M1}$, which are related through the domain matrix $[\boldsymbol{K}_{M1}]$ = $\begin{pmatrix} 0 & 0 & 2 \\ 0 & 1 & 0 \\ -1 & 0 & -1 \end{pmatrix} [\boldsymbol{K}_R]$. The diffraction intensity $I(\boldsymbol{K})$ is proportional to the square of the crystal structure factor $|F(\boldsymbol{K})|^2$. From the reported crystal structure [22], we calculated the $I_{cal}$(-12, 1, 1)$_R$ = 7.015 and $I_{cal}$(2, 1, 11)$_{M1}$ = 0.7676. The unit cell volume of the M1 structure is approximately twice that of the R phase ($V_{M1} \simeq 2 \times V_R$). It is also assumed that in the M1 phase, four types of domains corresponding to the tetragonal-to-monoclinic transition exist in nearly equal proportions. Therefore, the scale factor $S$ for calculating the ratio of the R and M1 domains is given by $S$ = [2$^2$ × $I_{cal}$(-12, 1, 1)$_R$] / [4 × $I_{cal}$(2, 1, 11)$_{M1}$] ~ 9.139. Using this $S$ value, the ratio of the R domains to the total is calculated as $I_{obs}$(-12, 1, 1)$_R$ / [$I_{obs}$(2, 1, 11)$_{M1}$ × $S$ + $I_{obs}$(-12, 1, 1)$_R$] (Fig. S5).

Figure S6 shows the current density dependence of the monoclinic lattice parameters of the M1 and R structures at $T_0$ = 300 K, where the lattice parameters of the R structure were converted using the following formula: $\boldsymbol{a}_m = 2\boldsymbol{c}_t$, $\boldsymbol{b}_m = \boldsymbol{b}_t$, and $\boldsymbol{c}_m = -\boldsymbol{a}_t - \boldsymbol{c}_t$.

Figure S7 shows the XRD data at various temperatures and current densities. Regardless of the current density, the diffuse scattering intensity is strong above $T_C$ = 340 K but becomes weak below $T_C$. At 300 K, the diffuse scattering remains weak at $J$ = 9 and 12 A/cm$^2$ (Fig. S7C), where the low-$J$ and high-$J$ structures are present, respectively, suggesting that the short-range correlation between V atoms is weaker than in the high-temperature phase.

Figure S8 shows the position dependence of the Raman spectra of VO$_2$ under an applied current of 15.00 A/cm$^2$. The incident light spot size is approximately 2 μm. Five different spectra were observed, with spectra of the same shape corresponding to dots of the same color in Fig. S8A. Figure S9 shows the current density dependence of the Raman spectra, which were obtained from a different measurement cycle than the one shown in Fig. S8. No clear dependence on current density was observed in the spectra corresponding to the M1 and R structures.

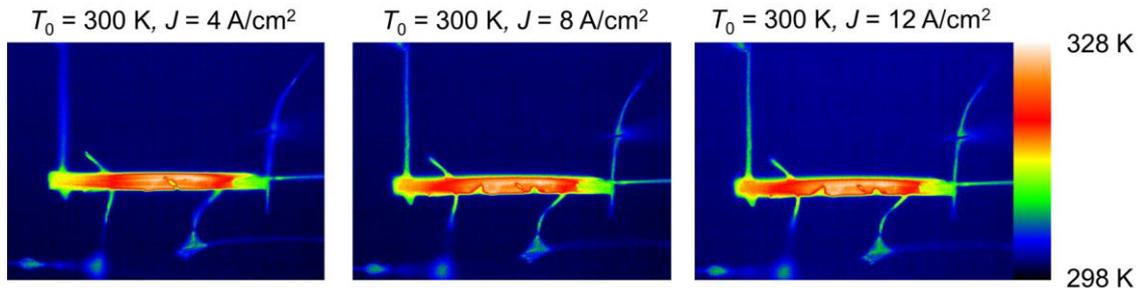

**Fig. S1.** Current density dependence of thermographic images of a single crystal of VO$_2$ at $T_0$ = 300 K.

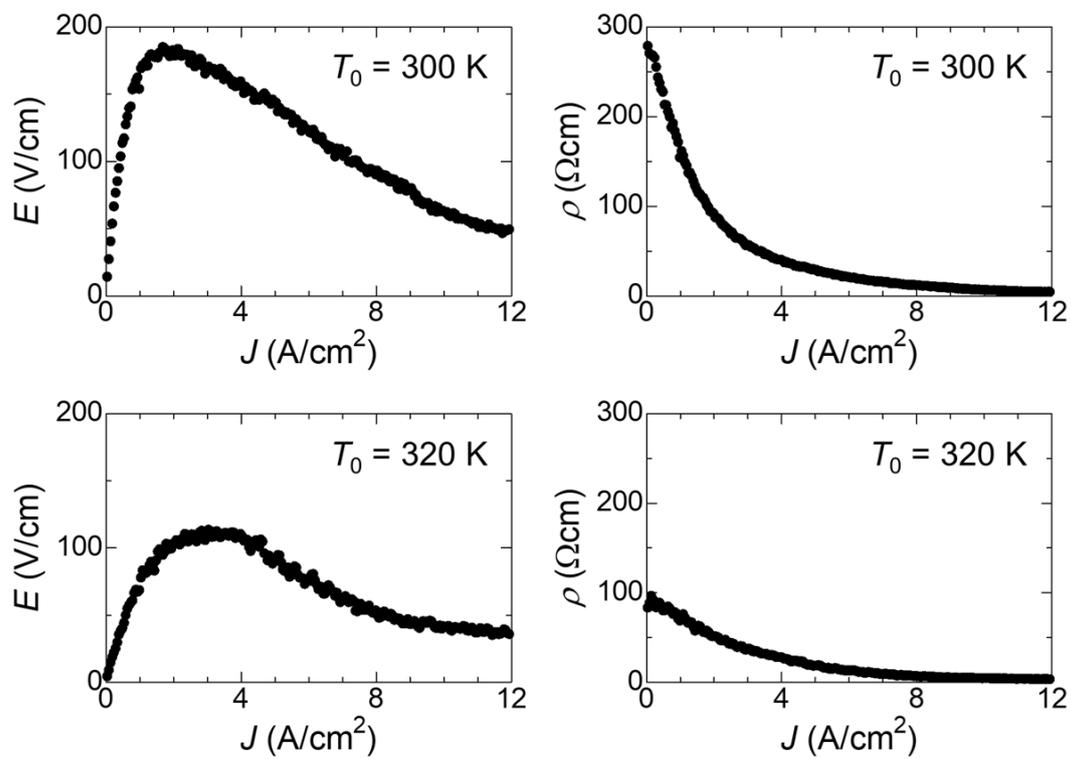

**Fig. S2.** Current density $J$ dependence of electric fields $E$ and resistivities $\rho$ at $T_0$ = 300 and 320 K, which were collected simultaneously with the XRD experiments.

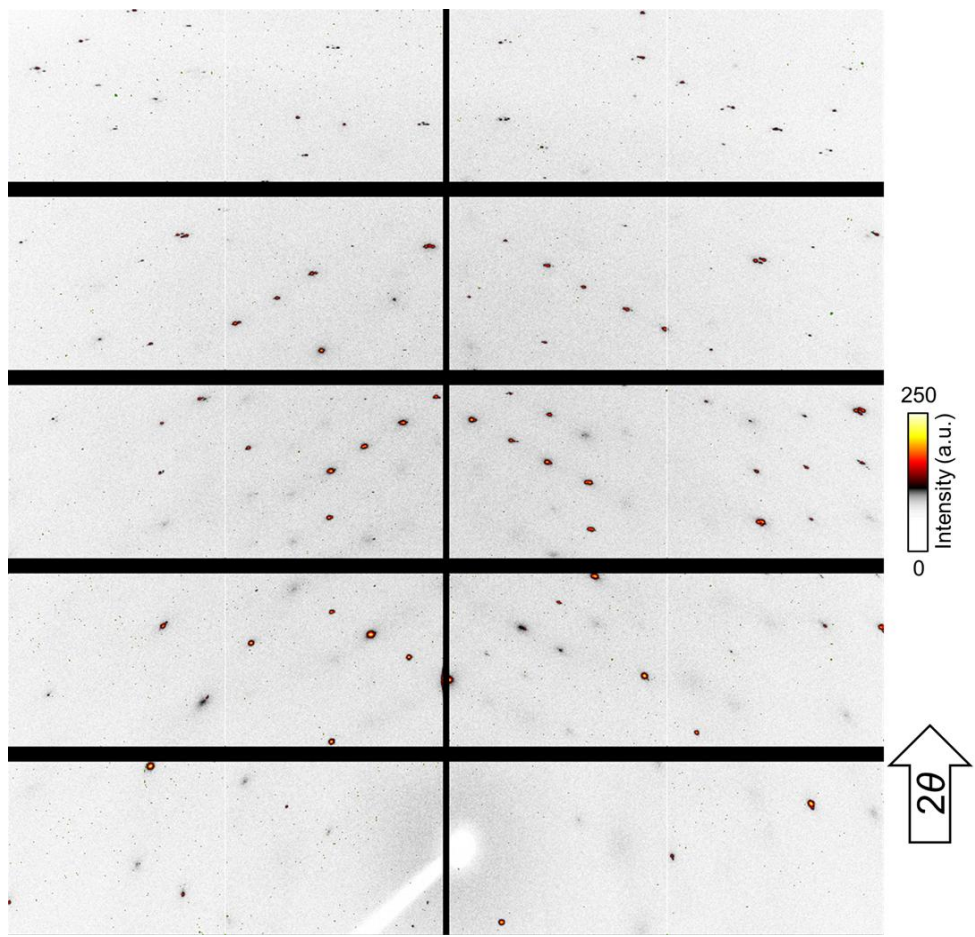

**Fig. S3.** Whole XRD image at $T_0$ = 300 K and $J$ = 0 A/cm$^2$. The data corresponds to a frame with $\Delta\omega$ = 1° within the measured range of $\omega$ = 0 to 180°. At the largest scattering angle, the resolution limit corresponds to $d_{min}$ = 0.28 Å.

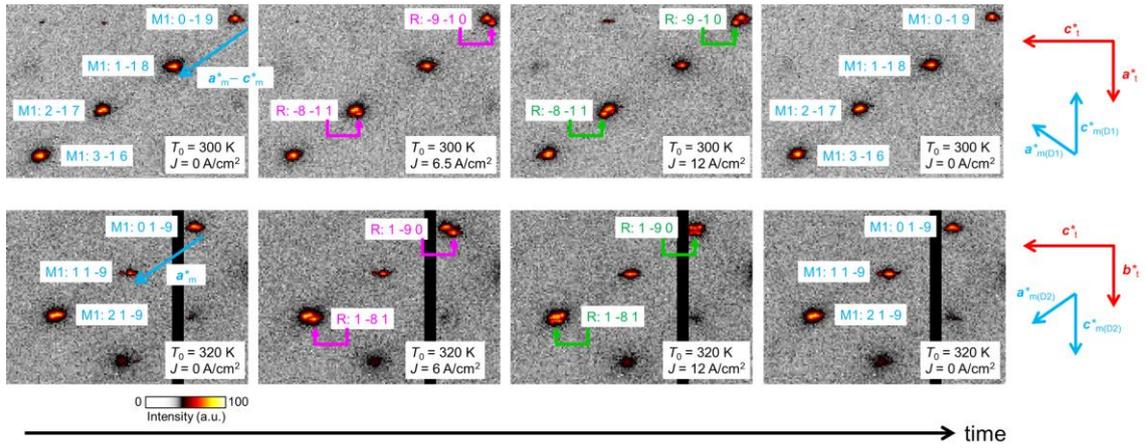

**Fig. S4.** Current density dependence of the XRD data at $T_0 = 300$ and 320 K is displayed in chronological order.

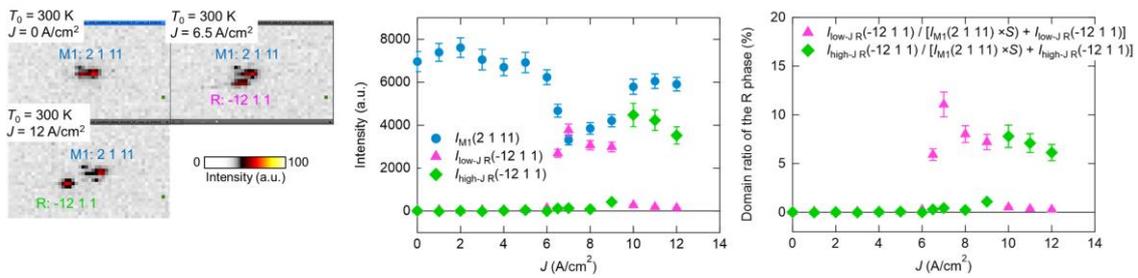

**Fig. S5.** Current density dependence of the domain ratio of the R phase at $T_0 = 300$ K.

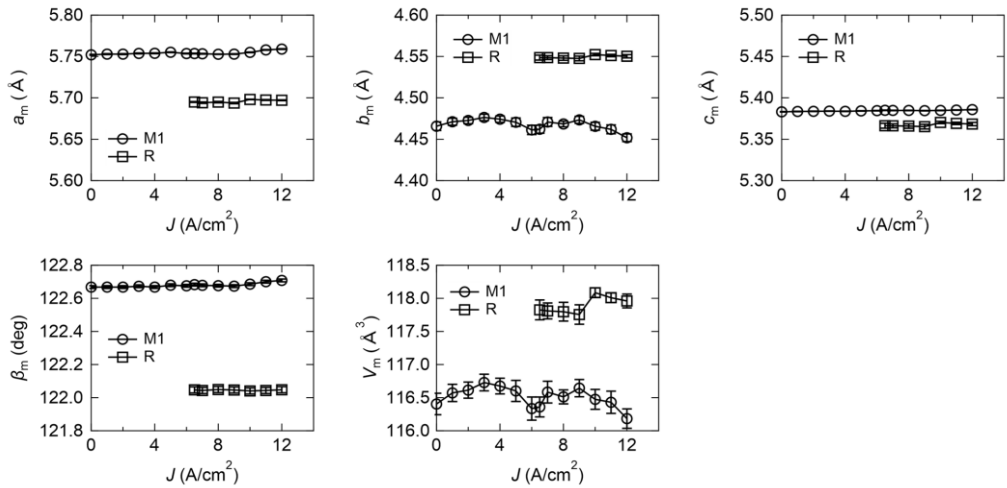

**Fig. S6.** Current density dependence of the monoclinic lattice parameters of the M1 and R structures at $T_0 = 300$ K.

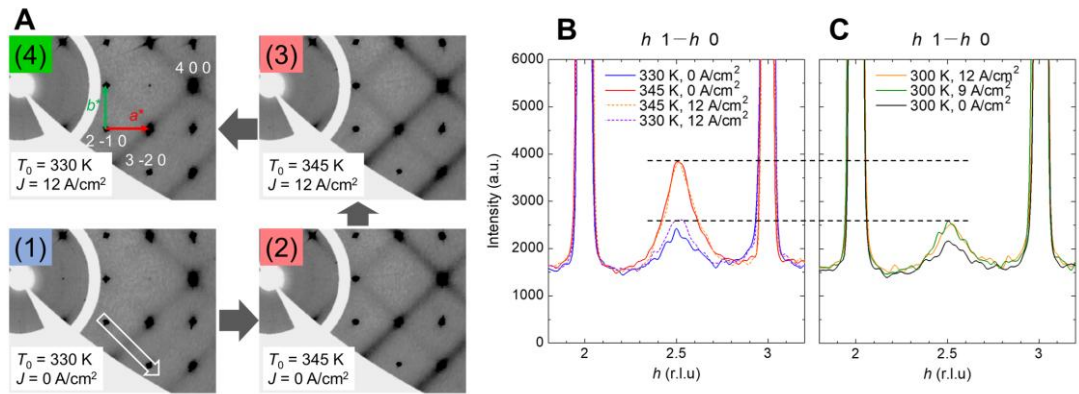

**Fig. S7.** (A) XRD data at various temperatures $T_0$ and current densities $J$. The numbers in parentheses indicate the measurement order. (B) One-dimensional plots of XRD intensity along the $h$ 1-$h$ 0 line corresponding to the data shown in (A). (C) $J$-dependence of the one-dimensional plots at $T_0$ = 300 K.

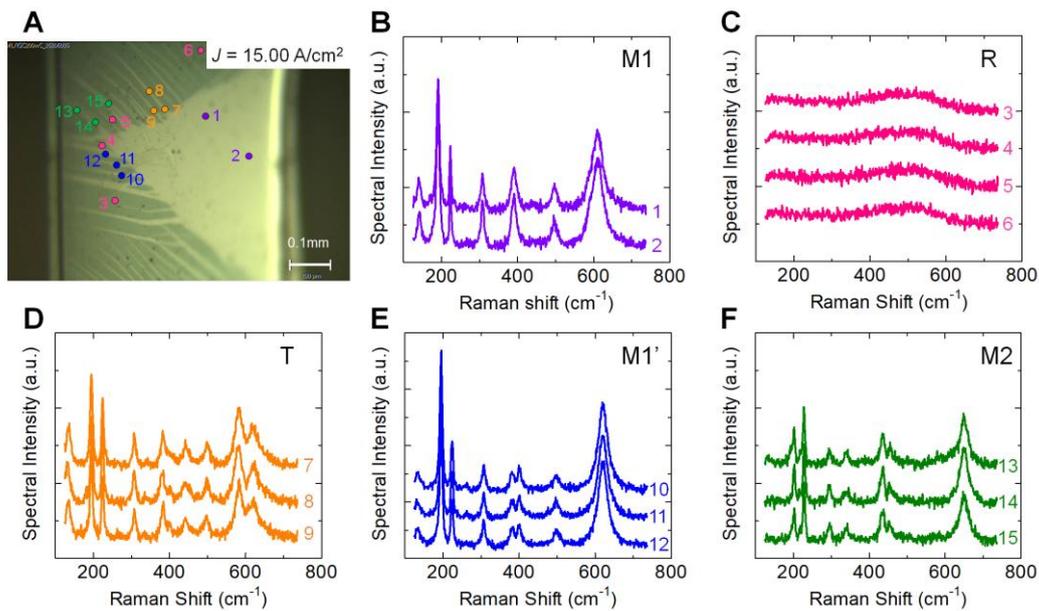

**Fig. S8.** (A) An optical image and (B-F) position dependence of the Raman spectra of $VO_2$ under an applied current of 15.00 A/cm². Each spectrum, labeled 1-15 in (B-F), corresponds to the positions 1-15 shown in (A). Purple, pink, orange, blue, and green spectra correspond to the M1, R, T, M1', and M2 structures, respectively.

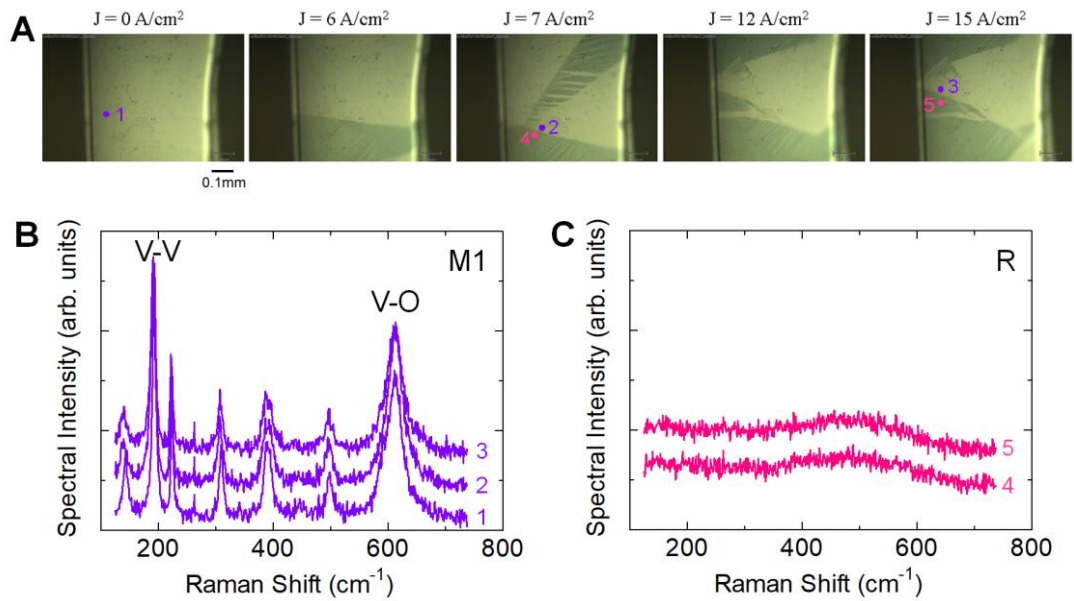

**Fig. S9.** Current density dependence of the (A) optical images and (B,C) Raman spectra. Each spectrum, labeled 1-5 in (B,C), corresponds to the positions 1-5 shown in (A).